\documentclass[pss]{wiley2sp} 
\usepackage{amsmath}

\begin{document}

\title{Low temperature magnetization of the quantum critical heavy fermion superconductor $\beta$-YbAlB$_4$}

\titlerunning{Low temperature magnetization of the heavy fermion system $\beta$-YbAlB$_4$}

\author{%
  Yosuke Matsumoto\textsuperscript{\Ast},
  Kentaro Kuga,
  Yoshitomo Karaki,
  Takahiro Tomita, and 
  Satoru Nakatsuji}

\authorrunning{Y. Matsumoto et al.}

\mail{e-mail
  \textsf{matsumoto@issp.u-tokyo.ac.jp}, Phone:
  +81-4-7136-3242, Fax: +81-4-7136-3242}

\institute{%
  Institute for Solid State Physics, Univ. of Tokyo, Kashiwa 277-8581, Japan.
  }

\received{XXXX, revised XXXX, accepted XXXX} 
\published{XXXX} 

\pacs{ } 

\abstract{%
%
%
%
\abstcol{%
$\beta$-YbAlB$_4$ is the first Yb-based heavy fermion superconductor with $T_{\rm c} = 80$ mK. 
We measured low temperature magnetization of  
high-purity single crystals down to $T=$ 25 mK. 
The measurements have revealed 
a considerable amount of volume fractions of the superconductivity and 
the upper critical field $B_{c2}$ curve under field along the $c$ axis,  
consistent with the previous results. 
  }{%
In the normal state, 
the previously observed divergent behavior in the temperature dependence of the magnetization has been confirmed using higher quality samples and under a low field of 22 mT.
In addition, the measurements have revealed a 
power law behavior, namely, $dM/dT\propto T^{3/2}$, 
which has a slightly higher exponent than 
the previous results.
	}}

%
%

\maketitle   

\section{Introduction}
Heavy fermion (HF) systems have attracted much interest as prototypical systems to study 
unconventional superconductivity and non-Fermi-liquid (NFL) states \cite{Mathur,Stewart,Yuan,Vojta,Lonzarich,Steglich}. 
In fact, a number of unconventional superconductors have been found 
in Ce (4$f^1$) based intermetallic HF systems near a quantum critical point 
\cite{Mathur,Yuan,Lonzarich}.
On the other hand, our recent studies have revealed 
the first Yb (4$f^{13}$) based HF superconductivity 
with the transition temperature $T_{\rm c}$ = 80 mK 
in the new compound $\beta$-YbAlB$_4$ \cite{beta2,beta3}. 
This is a strongly type-II
heavy fermion superconductor in the clean limit.
Upper critical fields are anisotropic, and strongly suppressed for the field along the $c$-axis, 
possibly because of the paramagnetic effect due to the divergent $c$-axis susceptibility. 
Strong sensitivity of $T_{\rm c}$ to sample purity suggests that the superconductivity 
is of an unconventional, non-$s$-wave type.

Interestingly, pronounced NFL behaviors appear above $T_c$ in 
the transport and thermodynamic properties \cite{beta2}. 
Furthermore, the magnetic field dependence of the NFL behaviors indicates that 
the system is a rare example of a pure metal that displays quantum criticality 
without external tuning, i.e. without doping, applied pressure and magnetic field \cite{beta2}. 

The Fermi surfaces (FS) topology has been also studied by quantum oscillation measurements \cite{beta4}. 
It has revealed three dimensional multi-FS of $f$-electrons which are roughly consistent with 
the band calculations \cite{beta4,beta5}. 
The bulk-sensitive hard x-ray photoemission spectroscopy measurements \cite{beta6} 
have clarified an intermediate Yb-valence of 2.745. 
Whether the valence fluctuation is relevant for the mechanism of the quantum criticality and superconductivity  is an interesting open question.  

In order to study the quantum criticality, 
detailed measurements of the basic thermodynamics are important. 
Here, we present the results of the low temperature magnetization measurements using a high sensitive SQUID magnetometer
and high-purity single crystals. 
Careful surface treatments of the single crystals and an improved stability of the measurements enabled us to observe the intrinsic behavior. 
We have confirmed the previously reported results, namely, a considerable amount of volume fractions for the superconductivity and 
the upper critical field $B_{c2}$ curve under a field along the $c$ axis.
In the normal state, the temperature dependence of the magnetization exhibits a divergent behavior 
down to $T_{\rm c}$ in a low magnetic field of 22 mT, which is consistent with the zero-field quantum criticality as previously reported \cite{beta2}. 
In addition, the measurements have revealed a power law behavior of $dM/dT\propto T^{3/2}$, which have a slightly larger exponent in comparison with the previous results.

\section{Experimental}
Single crystals of $\beta$-YbAlB$_4$ were prepared using aluminum self flux method \cite{beta1}. 
Most of single crystals form in a thin plate shape. 
Typical size used for the measurements is 
$\sim 0.5$ mm$^{2}$ $\parallel$ $ab$-plane $\times 0.01$\ mm $\parallel$ $c$-axis 
and has a smaller surface area along the plane than the one in the previous work \cite{beta3}. 
Surface impurities were removed carefully with dilute nitric acid before the measurements.
High-purity single crystals with RRR $>$ 200 ($\sim$ 30 pieces, 0.82 mg) were 
stacked using silver paste. This assembly was inserted in a pick-up coil and cooled by a heat link made of silver foils.  
The superconducting magnet was covered with a Nb superconducting shield and a $\mu$-metal tube in order 
to eliminate the earth's magnetic field. 
The pick-up and primary lines covered with Pb superconducting tubes 
are connected to a multi-purpose dc-SQUID probe 
located in a bath of liquid $^4$He at 4.2 K. 
The magnetization $M$ and ac-susceptibility $\chi _{ac}$ were obtained as dc and ac signals of the SQUID output. 
For the $\chi _{ac}$ measurement, an ac field of 0.1 $\mu$T and the
frequency $f=$ 16 Hz is applied along the $c$-axis.
The residual magnetic field of 1.1 $\mu$T was determined by 
field cooled sequences of $M$ measurements and adjusted by applying an excess magnetic field. 
The absolute values of $M$ were calibrated by 
comparing the data with those measured by a commercial SQUID magnetometer above 2.0 K. 

\section{Results and discussion}
\begin{figure}[t]%
\centering
\includegraphics*[width=0.79\linewidth]{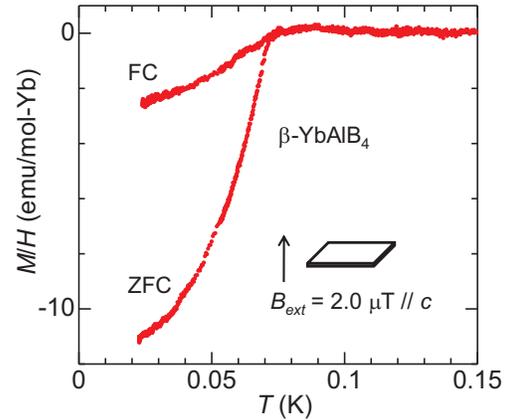}
\caption{%
Temperature dependence of the field-cooled (FC) and zero-field-cooled data (ZFC) 
of the dc-susceptibility $M/H$ under a field of 2.0 $\mu$T along the $c$-axis. 
}
\label{FC-ZFC}
\end{figure}

First, we present the temperature dependence 
of the dc-susceptibility $M/H$ measured under a small field along 
the $c$-axis (Fig.\ref{FC-ZFC}). 
Clear diamagnetic signals are observed in the susceptibility obtained 
through the field-cooled (FC) and the zero-field-cooled (ZFC) sequences. 
The signals are $\sim$ 1/3 of the previous results 
because of the smaller demagnetization effect due to the smaller 
cross section of the crystals normal to the field. 
Thus, 
the volume fractions are estimated to be of the same order 
as the previous results, i.e., 6\% and 40\% for the FC and ZFC respectively. 
The considerable amount of volume fractions indicates the bulk superconductivity \cite{beta3}.

\begin{figure}[t]%
\centering
\includegraphics*[width=0.95\linewidth]{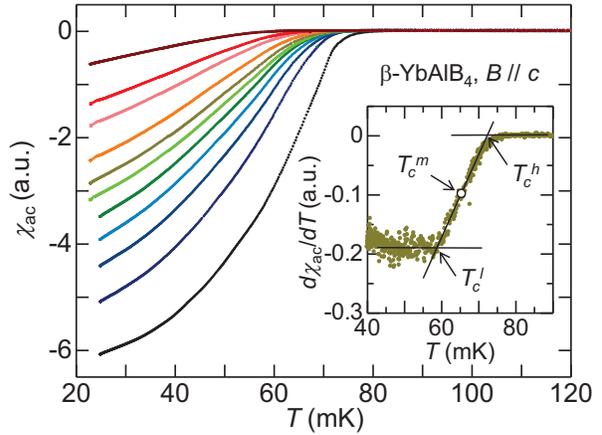}
\caption{%
Temperature dependence of the ac-susceptibility $\chi _{ac}$ under fields along the $c$-axis. The 
applied dc fields are 0, 0.31, 0.62, 0.93, 1.2, 1.6, 1.9, 2.5, 3.7, 5.0, 9.3 mT from bottom to top. 
Inset: $d\chi_{ac}/dT$ vs $T$ under $B$ = 1.9 mT. Three temperatures are defined by the 
points of intersections of the solid lines and their mid-point (open circle)(see text). 
}
\label{chiACvsT}
\end{figure}

The temperature dependence of $\chi _{ac}$ under fields along the $c$-axis 
is shown in Fig. \ref{chiACvsT}. 
The data shows clear diamagnetic signals and 
does not saturate down to the lowest $T$ of 25 mK even at $B = 0$. 
With increasing $B$, $T_c$ shifts to a lower $T$ and 
the diamagnetic signals become smaller and broader. 
In order to determine $T_c$ without an ambiguity, 
we take a temperature derivative of the data, i.e., $d\chi_{ac}/dT$ vs $T$ as shown in the inset of 
Fig. \ref{chiACvsT}.
With decreasing $T$ across $T_c$, $d\chi_{ac}/dT$ data starts to deviate from a high $T$ value and 
shows a step like drop with a finite slope. 
The higher and lower edge temperatures of the step, which can be defined as intersection points of the solid lines in the inset of Fig. \ref{chiACvsT}, are named $T_c^h$ and $T_c^l$, respectively. 
We define $T_c^m$ as a mid point of these two temperatures.

\begin{figure}[t]%
\centering
\includegraphics*[width=0.90\linewidth]{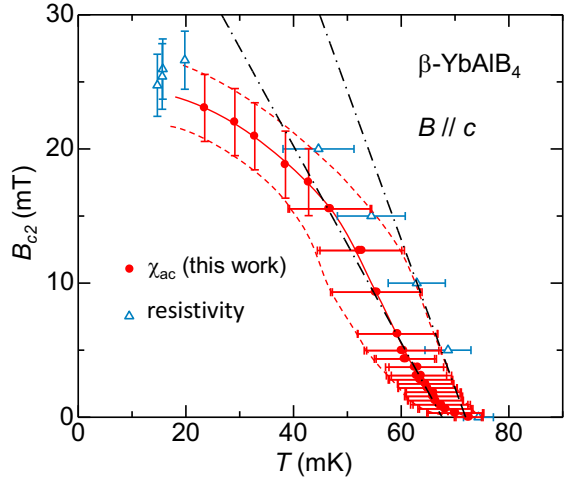}
\caption{%
Temperature dependence of the upper critical field $B_{\rm c2}$ along the $c$-axis 
determined by the $\chi _{ac}$ measurements. 
Open triangles are the data obtained by the previous resistivity measurements \cite{beta3}. 
Solid and broken lines are guides to the eye to show the field dependence of $T_c^l$, $T_c^m$ and $T_c^h$. 
The dot-dashed lines define $dB_{\rm c2}/dT$ for $T_c^m$ and $T_c^h$ (see text).
  }
\label{Bc2curve}
\end{figure}

Figure \ref{Bc2curve} presents the $B_{\rm c2}$ curve obtained by the $\chi _{ac}$ measurements. 
Here we plot $T_c^m$ by filled symbols, and $T_c^h$ and $T_c^l$ by 
the edges of the error bars. 
At $T < 45$ mK, we obtained the data by field sweep and 
determined $T_c$ by taking the field derivative of the susceptibility, 
adopting similar definitions to the case of the temperature sweep.  
The $B_{\rm c2}$ curve 
is consistent with the previous results of the resistivity measurements \cite{beta3}. 
In Fig. \ref{Bc2curve}, we plot $T_c$ by the resistivity measurements namely, 
the mid point by open triangles, the onset and zero resistivity temperatures by the edge of the error bars. 
Note that the mid points of resistivity agree well with $T_c^h$. 
The zero temperature value of 
an orbital critical field can be 
evaluated by using the Werthamer-Helfand-Hohenberg model 
as $B_{\rm c2}^{\rm orb}(0) = 0.727 (dB_{\rm c2}/dT)T_{\rm c}$. 
Because 
$B_{\rm c2}$ curve is convex below 2 mT possibly  
due to a distribution of $T_c$, we estimate 
$dB_{\rm c2}/dT$ at the field between 2 and 10 mT for $T_c^m$ and $T_c^h$ 
and obtain $B_{\rm c2}^{\rm orb}(0) =53$ mT and 80 mT for $T_c^m$ and $T_c^h$, respectively.
The observed $B_{\rm c2}(0)\sim$ 25 mT is much smaller than the estimates for $B_{\rm c2}^{\rm orb}(0)$ 
and the Pauli limit ($\sim$ 150 mT). This is most likely due 
to the paramagnetic depairing effect enhanced by the
large Ising magnetization along the $c$ axis \cite{beta3}.

\begin{figure}[t]%
\centering
\includegraphics*[width=0.95\linewidth]{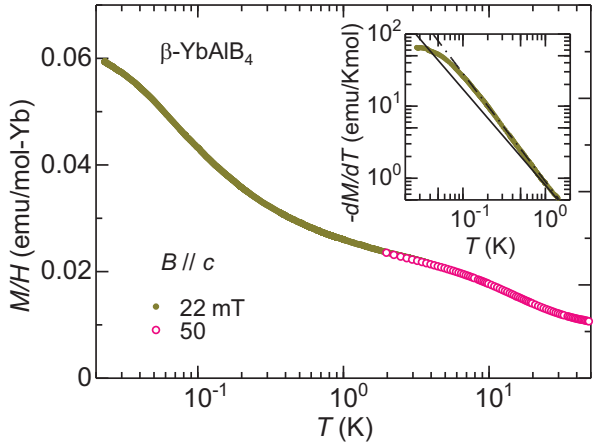}
\caption{%
Temperature dependence of the dc-susceptibility $M/H$ under fields along the $c$-axis. 
Inset: $-dM/dT$ vs $T$ at $B$ = 22 mT. The dot-dashed line and the solid lines represent 
$T^{-3/2}$ and $T^{-4/3}$ power law behaviors, respectively. 
  }
\label{MoverH}
\end{figure}

Temperature dependence of the dc-susceptibility $M/H$ under fields along the $c$-axis are 
shown in Fig. \ref{MoverH}. 
The data above $\sim$ 2 K show no field dependence below $B$ = 50 mT. 
Under the field of 22 mT, the superconducting signal becomes significantly suppressed owing to the suppression of $T_{\rm c}$ down to $\sim$ 28 mK. 
On the other hand, the divergent increase of $M$ in the normal state above $T_{\rm c}$ is consistent with 
the previous results, indicating the  quantum criticality of the system \cite{beta2}.
The inset of Fig. \ref{MoverH} shows the temperature derivative of $M(T)$, $dM/dT$. 
It exhibits a power law dependence of $T^{-3/2}$ rather than 
$T^{-4/3}$, which is expected from the previously reported asymptotic behavior, $M\propto T^{-1/3}$. 
However, the present data should represent the intrinsic behavior, because 
(1) only high quality single crystals were used, 
(2) a more detailed care was taken in the etching process 
to remove the possible sample surface impurities, and 
(3) the current data were obtained without a drift of SQUID output nor a sizable background signal \cite{beta3}. 
Note that the higher $T$ data obtained by a commercial SQUID magnetometer are always consistent 
with each other after the careful surface treatment. 
In addition, in the current measurements, the stability of the 
measurements was improved and the error coming from the drift and the background are carefully examined. 

To conclude, our low temperature magnetization measurements have revealed 
the asymptotic power law dependence 
$dM/dT\propto T^{-3/2}$ as an 
intrinsic low $T$ behavior. In addition, we have confirmed 
a considerable amount of volume fractions of the superconductivity and 
the $B_{c2}$ curve under a field along the $c$-axis 
which are all consistent with the previous results \cite{beta3}. 
The detailed field dependence of the magnetization to examine the nature of the quantum criticality and 
higher $T$ data for the precise crystal field analyses 
will be reported elsewhere \cite{beta7}.

\begin{acknowledgement}
We thank 
Y. Machida, L. Balicas, T. Sakakibara, H. Ishimoto for the support and useful discussions.
This work has been supported in part by
Grants-in-Aid for Scientific Research (21684019) from JSPS, Grant in 
Aid for Scientific Research on Priority Areas (17071003) from
MEXT of Japan.
\end{acknowledgement}

%

\begin{thebibliography}{[1]}
\bibitem{Mathur} 
N. D. Mathur {\it et al.}, Nature {\bf 394}, 39 (1998).

\bibitem{Stewart} 
G. R. Stewart, Rev. Mod. Phys. {\bf 73}, 797 (2001).

\bibitem{Yuan} 
H. Q. Yuan, F. M. Grosche, M. Deppe, C. Geibel, G. Sparn, and F. Steglich, Science {\bf302}, 2104 (2003).

\bibitem{Vojta} 
H. v. L\"{o}hneysen, A. Rosch, M. Vojta, and P. W\"{o}lfle, Rev. Mod. Phys. {\bf79}, 1015 (2007).

\bibitem{Lonzarich} 
P. Monthoux, D. Pines, and G. G. Lonzarich, Nature {\bf 450}, 1177 (2007).

\bibitem{Steglich} 
P. Gegenwart, Q. Si, and F. Steglich, Nature Phys. {\bf 4}, 186-197 (2008).

\bibitem{beta2} 
S. Nakatsuji {\it et al.}, 
Nature Phys. {\bf 4}, 603-607 (2008).

\bibitem{beta3} 
K. Kuga, Y. Karaki, Y. Matsumoto, Y. Machida, S. Nakatsuji, Phys. Rev. Lett. {\bf101}, 137004 (2008).

\bibitem{beta4} 
E. C. T. O'Farrell {\it et al.}, 
Phys. Rev. Lett. {\bf102}, 216402 (2009).

\bibitem{beta5}
A. H. Nevidomskyy and P. Coleman, Phys. Rev. Lett. {\bf102}, 077202 (2009).

\bibitem{beta6} 
M. Okawa {\it et al.}, 
arXiv:0906.4899. 

\bibitem{beta1} 
R. T. Macaluso {\it et al.}, 
Chem. Mater., {\bf19}, 1918 (2007). 

\bibitem{beta7} 
Y. Matsumoto {\it et al.}, {\it preprint}.



\end{thebibliography}
%

\end{document}